# Quantum Mechanics and Black Holes

Jose N. Pecina-Cruz
The University of Texas-Pan American
Department of Physics
1201 West University Drive,
Edinburg, Texas 78541
E-mail: jpecina2@panam.edu

**Abstract**

This paper discusses the existence of black holes from the foundations of quantum mechanics. It is found that quantum mechanics rule out a possible gravitational collapse.

PACS 11.30.-j, 11.30.Cp, 11.30.Er

**Introduction**

In a popular publication, Oppenheimer and Volkoff [1] suggested a gravitational collapse under the assumption of the validation of the statistics of Fermi-Dirac for an ensemble of neutrons separated by distances less than the neutron's Compton wavelength. Starting with the equation of state of a gas relativistic degenerated of Fermions (eqn. (11) in reference [1]), considering Einstein's equations, which are obtained from a line element exhibiting spherical symmetry. And removing the singularity at the Schwarzschild radius the final state equation was analyzed, for a neutron star with a mass beyond of that of gravitational equilibrium. A neutron star of $M_{max} > 0.76 \odot$ and $R_{min} = 9.42$ km collapses up to a radius = 0. However, Fermi-Dirac statistics is not applicable for distances shorter than the Compton wavelength of the particles. Heinsenberg's uncertainty principle [8] limits its range of validation. Landau and Lifshitz, in their exposition of the gravitational collapse [10], admitted that the physics they developed for a gravitational collapse is invalid for distances on the order of the Compton wavelength of an elementary particle. In the case of a Fermi gas of neutrons, this distance is small but greater than zero. Then, such collapse is not justified in a quantum theory of gravity. One more argument against the existence of black holes is the following: in particle physics is proved that the graviton is the exchange agent responsible for the transmission of the gravitational force. But, the graviton having a zero mass is similar to the photon. Both of the particles; photon and graviton are trapped by a black hole. Therefore, the existence of gravitation outside of a black hole rests on pure classical physics. This conclusion has also been reached in references [9],[10],[11], but not discussed in detail as this paper does. Section 1 is devoted to explain the physical conditions for the formation of antiparticles. Section 2 discusses the inconsistence of black holes with quantum mechanics.

## 1. Creation of Antiparticles

In the one particle scheme Feynman and Stückelberg interpret antiparticles as particles moving backward in time [3]. This argument is reinforced by S. Weinberg who realizes that antiparticles existence is a consequence of the violation of the principle of causality in quantum mechanics [2]. The temporal order of the events is distorted when a particle wanders in the neighborhood of the light cone. How is the antimatter generated from matter? According to Heisenberg's uncertainty principle a particle wandering in the neighborhood of the light-cone suddenly tunnels from the timelike region to the spacelike; in this region the relation of cause and effect collapses. Since an event, at $x_2$ is observed by an observer A, to occur later than one at $x_1$, in other words $x_2^0 > x_1^0$. An observer B moving with a velocity **v** respect to observer A, will see the events separated by a time interval given by

$$x'^0_2 - x'^0_1 = L^0_\alpha(v)(x_2^\alpha - x_1^\alpha), \qquad (1)$$

where $L^\beta_\alpha(v)$ is a Lorentz boost. From equation (1), it is found that if the order of the events is exchanged for the observer B, that is, $x'^0_2 < x'^0_1$ (the event at $x_1$ is observed later than the event at $x_2$.), then a particle that is emitted at $x_1$ and absorbed at $x_2$ as observed by A, it is observed by B as if it were absorbed at $x_2$, before the particle were emitted at $x_1$. The temporal order of the particle is inverted. This event is completely feasible in the neighborhood of the light-cone, since the uncertainty principle allows a particle tunnel from time-like to space-like cone regions. That is the uncertainty principle will consent to the space-like region reach values above than zero as is shown in next equation,

$$(x_1 - x_2)^2 - (x_1^0 - x_2^0)^2 \leq \left(\frac{\hbar}{mc}\right)^2,$$

where $\left(\frac{\hbar}{mc}\right)^2 > 0,$ \qquad (2)

and $\frac{\hbar}{mc}$ is the Compton wavelength of the particle. The left hand side of the equation (2) can be positive or space-like for distances less or equal than the square of the Compton wavelength of the particle. Therefore, causality is violated. The only way one can interpret this phenomenon is by assuming that the particle is absorbed at $x_2$, before it is emitted at $x_1$ as it is observed by B, is actually a particle with negative energy, mass, charge and certain spin, moving backward in time; that is $t_2 < t_1$ [3]. This event is equivalent to see an antiparticle moving forward in time with positive energy, mass, charge and opposite spin that it is emitted at $x_1$ and it is absorbed at $x_2$. With this reinterpretation the causality is recovered.

The square of the rest mass of a particle $p^2 = -m^2$ is an invariant of the Poincare group, but the rest mass is not an invariant [4][5]. Therefore, two observers on different inertial frames can see the same particle with different energy and mass signs. Observer A sees a particle of positive energy and mass moving forward in time, while observer B sees the same particle moving backward in time with negative energy and mass.

In this article is considered that a particle in the future timelike cone tunnel to the past timelike cone, that is from a positive to a negative energy. Since, in the spacelike region the momentum of particle is an imagery number. Equation 2 only guarantees a finite probability that a particle can tunnel from the future to the past timelike cone. The unitary irreducible representations of the "full" Poincare group (Poincare group with reflections) describe the elementary particles in the past timelike cone [11].

A particle of kinetic positive energy greater than $mc^2$ in the future timelike cone tunnels to the past timelike cone, through a barrier mainly given by the spacelike region, to acquire a negative kinetic energy less than $-mc^2$.

This section is concluded with the remark that the transition from positive energy to negative energy is an uphill event, since a particle has to have a velocity greater than the speed of light, according to equation 2. This event is classically impossible. However, quantum mechanics makes this event feasible by the Heisenberg's uncertainty principle. This violation of the second postulate of the special theory of relativity is only temporary and constrained by the uncertainty principle.

**2. Gravitational Collapse**

Negative energy states are significant for distances on the order of the Compton wavelength of a particle. According to Heisenberg's principle, if one tries to observe a particle with a very good resolution, one must perturb it with at least an energy-momentum equal to its rest mass. As a result of this perturbation oscillations of time could occur, generating antiparticles [7]. Therefore, before an assembly of particles, that collapses, reaches distances less than their Compton wavelengths creation of particles-antiparticles pairs occurs.

According to reference 1, if a cooling star does not reach the equilibrium as a white dwarf or a neutron star, that is, its mass during the thermonuclear evolution does not drop below the Chandrasekhar or Oppenheimer-Volkoff limits, will collapse reaching a state of infinite proper energy density in a finite time. But, the uncertainty principle given by equation (2), it rules out the gravitational collapse. A great activity of creation and annihilation of particles and antiparticles would take place when the separation between two particles is on the order of their Compton wavelength, unless an unknown nuclear reaction takes place. According to the authors of reference 1, pages 381-382, there only are two possible answers to the question of the "final" behavior of a very massive neutron star, either: 1) the equation of state fails or 2) the star collapses to form a black hole. The first answer is correct, since the particles of the Fermi statistics (equation of state) is not able to describe the physics for distances shorter than the Compton wavelength of the particles. Perhaps, a nuclear reaction (thermodynamic favorable) could take place before a possible gravitational collapse. Since such collapse would be quantum mechanically inconsistent because the uncertainty principle predicts the creation of particle-antiparticle (see equation. 2) pairs before any collapse take place. The quantum properties of black holes (if a collapse is feasible) were already examined by S. Hawking [8]. One should search in the universe burst of Gamma rays and Pions instead of the black holes. The concept of black hole is of pure classical nature ignoring the principles of quantum mechanics. Hawking has already pointed out that a theory of quantum gravity precludes the existence of these creatures [9].

**Conclusion**

This paper presented a logical scheme that showed the incompatibility of the physics of a collapsing star with the principles of quantum mechanics. This manuscript is intended to raise a rational doubt over the argument initiated by Oppenheimer and Volkoff in 1939 [1]. However, the concept of a black hole has reached the same level of popularity than that of the "ether cosmic." Black holes will still remain in the open for a long time.

**Acknowledgment**

I would like to thank Roger Maxim Pecina for his illuminating philosophical discussions. I am also in debt to Dr. Liang Zeng for encouraging me to publish this manuscript.